\documentclass[twocolumn,english,prb,amsmath,amssymb,showpacs]{revtex4}
\usepackage[T1]{fontenc}
\usepackage[latin9]{inputenc}
\usepackage{amsbsy}
\usepackage{graphicx}
\usepackage{amssymb}

\usepackage{babel}

\begin{document}

\title{Thermoelectric spin transfer in textured magnets}

\author{Alexey A. Kovalev}

\author{Yaroslav Tserkovnyak}

\affiliation{Department of Physics and Astronomy, University of California, Los
Angeles, California 90095, USA}

\date{\today}
\begin{abstract}
We study charge and energy transport in a quasi-1D magnetic wire in
the presence of magnetic textures. The energy flows can be expressed
in a fashion similar to charge currents, leading to new energy-current
induced spin torques. In analogy to charge currents, we can identify
two reciprocal effects: spin-transfer torque on the magnetic order
parameter induced by energy current and the Berry-phase gauge field
induced energy flow. In addition, we phenomenologically introduce
new $\beta-$like viscous coupling between magnetic dynamics and energy
current into the LLG equation, which originates from spin mistracking
of the magnetic order. We conclude that the new viscous term should
be important for the thermally induced domain wall motion. We study
the interplay between charge and energy currents and find that many
of the effects of texture motion on the charge currents can be replicated
with respect to energy currents. For example, the moving texture can
lead to energy flows which is an analogue of the electromotive force
in case of charge currents. We suggest a realization of cooling effect
by magnetic texture dynamics. 
\end{abstract}


\maketitle

The notion of the Berry phase\cite{Berry:mar1984} naturally appears
in the description of magnetic texture dynamics in the limit of strong
exchange field.\cite{Barnes:jun2007,Tserkovnyak:apr2008} The spin
up and down with respect to local magnetization electrons experience
{}``fictitious'' electromagnetic fields.\cite{Volovik:mar1987}
These fields have opposite signs for spin up and down electrons and
result in the Lorentz force.\cite{Barnes:jun2007} It has been realized
that the spin-transfer torque (STT) is a reciprocal effect to the
electromotive force (EMF) associated with this Lorentz force.\cite{Tserkovnyak:jan2009}
In real systems, the exchange field is finite leading to spin misalignments
with the texture, and more realistic description should take such
effects into account via $\beta$ terms in the Landau-Lifshitz-Gilbert
(LLG) equation.\cite{Zhang:sep2004,Tserkovnyak:apr2008a}

Recently, interest in thermoelectric effects has considerably increased
as new experimental data has been available.\cite{Sales:feb2002}
The Peltier effect describes heat transfer accompanying the current
flow. The opposite is the Seebeck effect that describes the thermo-EMF
induced by temperature gradients. The Peltier and Seebeck thermoelectric
effects as well as the thermoelectric STTs have been studied in multilayered
nanostructures.\cite{Hatami:may2009} Hatami \textit{et al.} proposed
a thermoelectric STT as mechanism for domain wall motion.\cite{Hatami:may2009}
Berger and Jen and Berger observed and discussed domain wall (DW)
motion induced by heat currents.\cite{Berger:jul1985} Thermal STTs
may soon be employed in the next generation of nonvolatile data elements
for reversal of magnetization. Thermoelectric nano-coolers can find
applications in the nanoelectronic circuits and devices.\cite{Ohta:feb2007} 

In this Rapid Communication, we study continuous magnetic systems
which can be relevant to DW motion\cite{Yamaguchi:Feb2004} and spin-textured
magnets.\cite{Muhlbauer:2009} We phenomenologically describe thermal
STTs in a quasi-1D magnetic wire with magnetic texture. The Berry-phase
gauge field induced energy flow turns out to be reciprocal effect
to the thermal STT and both effects can be formally eliminated from
the equations of motion by properly redefining the thermodynamic variables
which is reminiscent of the non-dissipative STTs.\cite{Tserkovnyak:jan2009}
We further generalize our description by including viscous $\beta-$like
terms corresponding to spin misalignments. These viscous effects turn
out to be important for the thermally induced DW motion and can lead
to such effects as cooling by magnetic texture dynamics. We also find
that the Peltier and Seebeck effects can be modified and tuned by
the magnetic texture dynamics.

\begin{figure}
\centerline{\includegraphics[bb=102bp 396bp 591bp 683bp,clip,width=0.8\columnwidth]{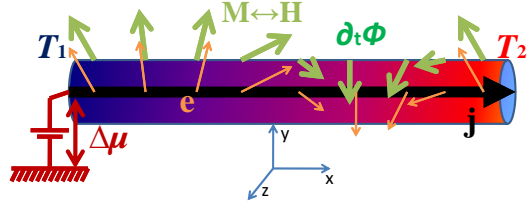}} 

\caption{(color online). In quasi-1D magnetic wire, charge current density
$j$ is induced by potential gradients $\partial_{x}\mu$, temperature
gradients $\partial_{x}T$ and EMF $\partial_{t}\Phi$ produced by
the Berry phase $\Phi$, which is acquired by the electron spin following
the instantaneous magnetic profile. Coupled viscous processes arise
once we relax the projection approximation. The magnetic texture $\mathbf{m}(x,t)$
responds to the effective field $\mathbf{H}(x,t)$.}

\label{fig} 
\end{figure}

Consider a thin quasi-1D magnetic wire with the magnetization $\mathbf{m}(x,t)$
(along the spin density) in the presence of chemical potential $\mu(x,t)$
and temperature $T(x,t)$ gradients. We would like to construct a
phenomenological description of our system based on thermodynamic
variables introduced above and their conjugate forces. The ferromagnetic
wire is supposed to be thermally isolated, and after being perturbed
by nonequilibrium chemical potential, magnetization and temperature
gradients, the wire evolves back toward equilibrium according to the
equations of motion, producing entropy. Note that we allow this equilibrium
state to be topologically nontrivial, \textit{e.g.} a magnetic DW
or vortex. The state of partial equilibrium can be described by thermodynamic
variables $x_{i}$ and their conjugates (generalized forces) $X_{i}=\partial\mathbb{S}/\partial x_{i}$
with the entropy and its time derivative being:\[
\mathbb{S}=\mathbb{S}_{0}-\dfrac{1}{2}{\displaystyle \sum_{i,k=1}^{n}\beta_{ik}x_{i}x_{k}},\:\dot{\mathbb{S}}=-{\displaystyle \sum_{i=1}^{n}X_{i}\dot{x}_{i}}\:.\]
We initially consider the entropy $\mathbb{S}(\rho,\rho_{U},\mathbf{m})$
as a function of the density of electron charge $\rho$, the density
of energy $\rho_{U}$ and the magnetization direction. The magnitude
of the magnetization is not treated as a dynamic variable, assuming
sufficiently fast spin-flip relaxation. The conservation laws of energy
and charge provide linear relations:\begin{equation}
\dot{\rho}=-\partial_{x}j\:,\:\dot{\rho}_{U}=-\partial_{x}j_{U}\:,\label{Conservation}\end{equation}
where we introduced the charge current $j$ and energy current $j_{U}$.
For conserved quantities, it is more convenient to work with fluxes
$j$ and $j_{U}$ instead of densities $\rho$ and $\rho_{U}$ which
leads to equivalent description due to the linear relations in Eqs.
(\ref{Conservation}). 

We will not be concerned with the general expression for the entropy
but rather concentrate on identifying the thermodynamic variables
and their conjugates by calculating the time derivative of the entropy.
Suppose that we fix the texture, then the rate of the entropy change
is:\cite{Landau:1984}\begin{equation}
\dot{\mathbb{S}}=-\oint dx\,\dfrac{\partial_{x}j_{U}+\mu\dot{\rho}}{T}=-\oint dx\,\left(\dfrac{\partial_{x}j_{q}}{T}+\dfrac{j\partial_{x}\mu}{T}\right),\label{Entropy_Rate}\end{equation}
where we introduced the modified energy current $j_{q}=j_{U}-\mu j$
that describes the energy flow without the energy corresponding to
the chemical potential $\mu$ (from now on only the energy current
$j_{q}$ is considered). The chemical potential is defined as a conjugate
of the density of charge. The introduction of $j_{q}$ is necessary
to avoid the unphysical gauge dependence of the energy current and
the associated kinetic coefficients on the potential offset for the
whole system. We are now ready to write the rate of the entropy change
for the general case of dynamic spin texture:\begin{equation}
\dot{\mathbb{S}}=\oint dx\,\partial_{x}\left(\dfrac{1}{T}\right)j_{q}-\oint dx\,\dfrac{\partial_{x}\mu}{T}j-\oint dx\,\dfrac{\mathbf{H}}{T}\cdot\partial_{t}\mathbf{m}\,,\label{Entropy_Rate_1}\end{equation}
where in Eq. (\ref{Entropy_Rate}) we integrated the term involving
$j_{q}$ by parts and the conjugate/force corresponding to the magnetization
is defined as $-\partial_{\mathbf{m}}\mathbb{S}|_{Q,q}=\mathbf{H}/T$
with $Q(x)$ and $q(x)$ being the overall charge and energy that
passed the cross section at point $x$ which corresponds to integrating
$j$ and $j_{q}$ in time, respectively. As it can be seen from Eq.
(\ref{Entropy_Rate_1}), our other conjugates are $-\partial_{q}\mathbb{S}|_{\mathbf{m},Q}=-\partial_{x}\left(1/T\right)$
and $-\partial_{Q}\mathbb{S}|_{\mathbf{m},q}=\partial_{x}\mu/T$.
In general, $\mathbf{H}$ is not the usual {}``effective field''
corresponding to the variation of the Landau free-energy functional
$F[\mathbf{m},\mu,T]$ and only when $\partial_{x}T=0$ and $\partial_{x}\mu=0$
the {}``effective fields'' coincide. Let us initially assume that
even in an out-of-equilibrium situation, when $\partial_{x}T\neq0$
and $\partial_{x}\mu\neq0$, $\mathbf{H}$ depends only on the instantaneous
texture $\mathbf{m}(x)$. In general, however, we may expand $\mathbf{H}$
phenomenologically in terms of small $\partial_{x}T$ and $\partial_{x}\mu$. 

In our phenomenological theory, the time derivatives of thermodynamic
variables are related to the thermodynamic conjugates via the kinetic
coefficients. In order to identify the kinetic coefficients, we assume
that the currents $j$ and $j_{q}$ are determined by the chemical
potential and temperature gradients as well as the magnetic wire dynamics,
which exerts fictitious Berry phase gauge fields\cite{Tserkovnyak:jan2009}
on the charge transport along the wire. We then have for the charge/energy
current gradient expansion:\begin{equation}
j=-\tilde{g}T\dfrac{\partial_{x}\mu}{T}+\tilde{\xi}T\dfrac{\partial_{x}T}{T^{2}}+\tilde{p}\left(\mathbf{m}\times\partial_{x}\mathbf{m}+\tilde{\beta}\partial_{x}\mathbf{m}\right)\cdot\partial_{t}\mathbf{m}\,,\label{ohmD}\end{equation}

\begin{equation}
j_{q}=\tilde{\xi}T\dfrac{\partial_{x}\mu}{T}-\tilde{\zeta}T^{2}\dfrac{\partial_{x}T}{T^{2}}+\tilde{p}'\left(\mathbf{m}\times\partial_{x}\mathbf{m}+\tilde{\beta}'\partial_{x}\mathbf{m}\right)\cdot\partial_{t}\mathbf{m}\,,\label{HeatCurr}\end{equation}
where we assume that the coefficients $\tilde{g}$, $\tilde{\xi}$
and $\tilde{\zeta}$ can in general also depend on temperature and
texture, for the latter, to the leading order, as $\tilde{g}=\tilde{g}_{0}+\eta_{\tilde{g}}(\partial_{x}\mathbf{m})^{2}$,
etc.. In Eqs. (\ref{ohmD}) and (\ref{HeatCurr}), we expand only
up to the linear order in the nonequilibrium quantities $\partial_{x}\mu$,
$\partial_{x}T$ and $\partial_{t}\mathbf{m}$ and to the second order
in $\partial_{x}\mathbf{m}$; however, the latter terms are expected
to be small in practice and only are necessary for establishing the
positive-definiteness of the response matrix. The spin-rotational
symmetry of the magnetic texture and the inversion symmetry of the
wire are also assumed to avoid additional and often complicated terms
in our expressions. Relating $\partial_{t}\mathbf{m}$ to the generalized
force $-\mathbf{H}/T$, within the LLG\cite{Gilbert:nov2004} phenomenology,
we derive the modified LLG equation consistent with Eqs. (\ref{ohmD})
and (\ref{HeatCurr}), with the guidance of the Onsager reciprocity
principle:\begin{equation}
\begin{array}{c}
s(1+\alpha\mathbf{m}\times)\partial_{t}\mathbf{m}+\mathbf{m}\times\mathbf{H}=-\tilde{p}\left[\partial_{x}\mathbf{m}+\tilde{\beta}(\mathbf{m}\times\partial_{x}\mathbf{m})\right]\\
\times\partial_{x}\mu-\tilde{p}'\left[\partial_{x}\mathbf{m}+\tilde{\beta}'(\mathbf{m}\times\partial_{x}\mathbf{m})\right]\cdot\partial_{x}T/T\,,\end{array}\label{LLGD-heat}\end{equation}
where we introduced the spin density $s$ so that $s\mathbf{m}=\mathbf{M}/\gamma$,
with $M$ being the magnetization density and $\gamma$ the gyromagnetic
ratio ($\gamma<0$ for electrons). Equation (\ref{LLGD-heat}) can
be expressed in terms of the charge/energy flows by inverting the
linear relation $\{j,j_{q}\}=\{\tilde{g},\tilde{\xi};\tilde{\xi},\tilde{\kappa}\}\{\partial_{x}\mu,\partial_{x}T/T\}$:

\begin{equation}
\partial_{x}\mu=-gj+\xi j_{q}+p\left(\mathbf{m}\times\partial_{x}\mathbf{m}+\beta\partial_{x}\mathbf{m}\right)\cdot\partial_{t}\mathbf{m}\,,\label{ohmD1}\end{equation}

\begin{equation}
\partial_{x}T/T=\xi j-\zeta j_{q}+p'\left(\mathbf{m}\times\partial_{x}\mathbf{m}+\beta'\partial_{x}\mathbf{m}\right)\cdot\partial_{t}\mathbf{m}\,,\label{HeatCurr1}\end{equation}

\begin{equation}
\begin{array}{c}
s(1+\alpha\mathbf{m}\times)\partial_{t}\mathbf{m}+\mathbf{m}\times\mathbf{H}=p\left[\partial_{x}\mathbf{m}+\beta(\mathbf{m}\times\partial_{x}\mathbf{m})\right]j\\
\qquad\qquad\qquad\qquad\qquad+p'\left[\partial_{x}\mathbf{m}+\beta'(\mathbf{m}\times\partial_{x}\mathbf{m})\right]j_{q}\,,\end{array}\label{LLGD-heat1}\end{equation}
where the new coefficients $g$, $\xi$, $\zeta$, $p$, $p'$, $\beta$
and $\beta'$ can be expressed via the ones marked by tilde, and in
Eq. (\ref{LLGD-heat1}), we disregarded the terms of the order $\sim(\partial_{x}\mathbf{m})^{2}\partial_{t}\mathbf{m}$
contributing to the Gilbert damping. Terms of similar order can also
appear in case of incompressible charge flow and lead to the non-local
Gilbert damping.\cite{Tserkovnyak:jan2009} The kinetic coefficients
contain information about the conductivity, $\sigma=\tilde{g}$, the
thermal conductivity, $\kappa=1/(\zeta T)$, and the conventional
Seebeck and Peltier coefficients can be found from Eqs. (\ref{ohmD})
and (\ref{HeatCurr1}) by assuming $j=0$ for the former, $S=-\tilde{\xi}/(\tilde{g}T)$,
and by assuming $\partial_{x}T=0$ for the latter, $\Pi=\xi/\zeta=-\tilde{\xi}/\tilde{g}$,
which also implies that $g=1/\sigma+S^{2}T/\kappa$. Equation (\ref{LLGD-heat1})
differs from an ordinary LLG equation\cite{Tserkovnyak:apr2008a}
by the extra spin torque terms that appear in the presence of the
energy flow $j_{q}$. These new torques are similar to the nondissipative
and dissipative current induced spin torques,\cite{Tserkovnyak:jan2009}
as the former can be related to electron spins following the magnetic
texture and the latter - to electron spins mistracking the texture.
The phenomenological parameter $\tilde{p}$ (or $p=\tilde{p}/\sigma_{0}-p'\Pi_{0}$)
can be approximated as $\tilde{p}/\sigma_{0}=\wp\hbar/2e$ in the
strong exchange limit\cite{Tserkovnyak:jan2009} and corresponds to
the electron spin-charge conversion factor $\hbar/2e$ multiplied
by the polarization $\wp=(\sigma_{0}^{\uparrow}-\sigma_{0}^{\downarrow})/\sigma_{0}$,
$\sigma_{0}=\sigma_{0}^{\uparrow}+\sigma_{0}^{\downarrow}$ and $e$
is minus the charge of particles, e.g. for electrons $e$ is positive. 

Similarly, we can consider Eq. (\ref{LLGD-heat1}) under conditions
of vanishing charge currents and fixed texture, and find the spin
current resulting from the temperature gradients: $2eS_{s}\partial_{x}T/(1/\sigma_{0}^{\uparrow}+1/\sigma_{0}^{\downarrow})$
where $S_{s}=(S_{0}^{\uparrow}-S_{0}^{\downarrow})/e$ is the spin
Seebeck coefficient in the absence of magnetic texture. By involving
the electron spin-charge conversion factor again, we can approximate
the second spin torque parameter $p'$ in Eq. (\ref{LLGD-heat1})
in the strong exchange limit arriving at\begin{equation}
p'=-\dfrac{\hbar}{2e}\wp_{S}S_{0}\dfrac{\sigma_{0}(1-\wp^{2})}{\kappa_{0}}\,,\, p=\dfrac{\wp\hbar}{2e}-p'\Pi_{0}\,,\label{ThermalPolarization}\end{equation}
where we introduced the spin polarization of the Seebeck coefficient
$\wp_{S}=(S_{0}^{\uparrow}-S_{0}^{\downarrow})/(S_{0}^{\uparrow}+S_{0}^{\downarrow})=eS_{s}/(2S_{0})$.
When the thermal conductivity is mostly due to electron motion, we
can simplify Eq. (\ref{ThermalPolarization}) further with the help
of the Wiedemann-Franz Law according to which $\kappa_{0}/\sigma_{0}=LT$
where $L=\pi^{2}k_{B}^{2}/(3e^{2})$ is the Lorenz number. Effects
such as spin drag\cite{D'amico:aug2000} can also influence the estimate
in Eq. (\ref{ThermalPolarization}). Note that the result in Eq. (\ref{ThermalPolarization})
can also be obtained from Eq. (\ref{HeatCurr1}) by considering the
texture-dynamics induced EMF which can lead to the energy currents
in the absence of charge currents. 

One can redefine the magnetization (which in turn leads to changes
in generalized forces $\partial_{x}\mu/T$ and $\partial_{x}T/T^{2}$)
in Eq. (\ref{Entropy_Rate_1}) so that the nondissipative parts of
torques are absorbed into these new definitions:\[
\begin{array}{c}
\partial_{t}\mathbf{\widetilde{m}}\rightarrow\partial_{t}\mathbf{m}-p\dfrac{1-\alpha\mathbf{m}\times}{(1+\alpha^{2})s}\partial_{x}\mathbf{m}\, j-p'\dfrac{1-\alpha\mathbf{m}\times}{(1+\alpha^{2})s}\partial_{x}\mathbf{m}\, j_{q}\,,\\
\partial_{Q}\mathbb{S}|_{\mathbf{\widetilde{m}}q}=\dfrac{1}{T}\left[\partial_{x}\mu-p(\mathbf{m}\times\partial_{x}\mathbf{m})\cdot\partial_{t}\mathbf{m}\right]\,,\\
\partial_{q}\mathbb{S}|_{\mathbf{\widetilde{m}}Q}=\dfrac{1}{T}\left[\dfrac{\partial_{x}T}{T}-p'(\mathbf{m}\times\partial_{x}\mathbf{m})\cdot\partial_{t}\mathbf{m}\right]\,,\end{array}\]
where with this choice of thermodynamic variables, Eqs. (\ref{ohmD1}),
(\ref{HeatCurr1}) and (\ref{LLGD-heat1}) will only have spin torque
terms proportional to $\beta(\beta')$. 

From Eq. (\ref{Entropy_Rate_1}), we can write the rate of the entropy
production:

\begin{equation}
\begin{array}{c}
\dot{\mathbb{S}}=\oint\dfrac{dx}{T}\left[gj^{2}+\zeta j_{q}^{2}-2\xi jj_{q}+\alpha s(\partial_{t}\mathbf{m})^{2}\right.\\
\left.\qquad-2\beta pj\partial_{x}\mathbf{m}\cdot\partial_{t}\mathbf{m}-2\beta'p'j_{q}\partial_{x}\mathbf{m}\cdot\partial_{t}\mathbf{m}\right]\,,\end{array}\label{Entropy_Rate_2}\end{equation}
where $g=g_{0}+\eta_{g}(\partial_{x}\mathbf{m})^{2}$, $\xi=\xi_{0}+\eta_{\xi}(\partial_{x}\mathbf{m})^{2}$
and $\zeta=\zeta_{0}+\eta_{\zeta}(\partial_{x}\mathbf{m})^{2}$. Notice
that the STTs in Eq. (\ref{LLGD-heat1}) induced by the charge/energy
currents can be separated into the nondissipative and dissipative
parts ($\beta$ terms) based on Eq. (\ref{Entropy_Rate_2}). This
separation is, nevertheless, formal as in realistic metallic systems
the torques will always be accompanied by the dissipation due to the
finite thermal conductivity $\kappa$ and conductivity $\sigma$.
The dissipation in Eq. (\ref{Entropy_Rate_2}) is guaranteed to be
positive-definite if the following formal inequalities hold:\begin{equation}
\begin{array}{c}
\xi\leq\sqrt{g\zeta},\;\eta_{g}\geq\dfrac{\beta^{2}p^{2}}{\alpha s},\;\eta_{\zeta}\geq\dfrac{\beta'^{2}p'^{2}}{\alpha s}\,,\\
\left(\eta_{\xi}-\dfrac{\beta p\beta'p'}{\alpha s}\right)^{2}\leq\left(\eta_{g}-\dfrac{\beta^{2}p^{2}}{\alpha s}\right)\left(\eta_{\zeta}-\dfrac{\beta'^{2}p'^{2}}{\alpha s}\right)\,,\end{array}\label{Inequalities}\end{equation}
where the first inequality can be rewritten equivalently as $g\geq S^{2}T/\kappa$,
and should always hold. Other inequalities are somewhat formal since
their proof implies that our theory can describe sufficiently sharp
and fast texture dynamics for dominating dissipation as opposed to
the first inequality for proof of which a mere static texture assumption
is sufficient. Nevertheless, Eqs. (\ref{Inequalities}) can serve
for estimates of the spin-texture resistivities ($\eta_{\zeta}$ and
$\eta_{g}$) and the spin-texture Seebeck effect ($\eta_{\xi}$) due
to spin dephasing. The condition on the spin-texture resistivity\cite{Tserkovnyak:jan2009}
$\eta_{1/\sigma}\geq(\tilde{\beta}\tilde{p})^{2}/(\alpha s\sigma_{0}^{2})$
also follows from Eqs. (\ref{Inequalities}). 

Let us now discuss thermal effects that can arise from the presence
of magnetic texture. In case of a static spin texture and stationary
charge density, $\partial_{x}j=0$, we can write the modification
to the Thomson effect by calculating the rate of heat generation\cite{Landau:1984}
$\dot{\mathbb{Q}}=-\partial_{x}j_{U}$ from Eqs. (\ref{ohmD}) and
(\ref{HeatCurr1}):\begin{equation}
\begin{array}{c}
\dot{\mathbb{Q}}=\kappa\partial_{x}^{2}T+(\partial_{T}\kappa)(\partial_{x}T)^{2}+\sigma j^{2}+T(\partial_{T}S)j\partial_{x}T\\
+\eta_{\kappa}\partial_{x}[(\partial_{x}\mathbf{m})^{2}]\partial_{x}T+T\eta_{S}\partial_{x}[(\partial_{x}\mathbf{m})^{2}]j\,,\end{array}\label{Thomson_Effect}\end{equation}
where $\kappa=\kappa_{0}+\eta_{\kappa}(\partial_{x}\mathbf{m})^{2}$
and $S=S_{0}+\eta_{S}(\partial_{x}\mathbf{m})^{2}$. Even though Eqs.
(\ref{ohmD}) and (\ref{HeatCurr1}) are the first order gradient
expansions, they are sufficient for obtaining Eq. (\ref{Thomson_Effect})
since consideration of the second order expansions would only lead
to even higher order terms in Eq. (\ref{Thomson_Effect}). The Thomson
effect can be modified by the presence of texture and some analogue
of local cooling may be possible even without temperature gradients.
However, the magnitude of the coefficients $\eta_{\kappa}$ and $\eta_{S}$
is not accessible at the moment and should be extracted from the microscopic
calculations. 

Another thermal effect we would like to discuss is related to heat
flows induced by magnetization dynamics. As can be seen from Eq. (\ref{HeatCurr1}),
such heat flows can appear even in the absence of temperature gradients
and charge current flows. When the magnetic texture follows a periodic
motion, the energy flows should result in effective cooling or heating
of some regions by specifically engineering the magnetic state of
the wire and applied rf magnetic fields. The spin spring magnets can
be of relevance.\cite{Uzdin:jun2008} Alternatively, the texture in
our wire (\textit{i.e.} spiral), can result from the Dzyaloshinskii-Moriya
(DM) interaction\cite{Landau:1984} relevant for such materials as
MnSi, (Fe,Co)Si or FeGe. In this case, the end of the spiral can be
exchange coupled to a homogeneous magnetization of a magnetic film
subject to rf magnetic field which should result in rotation of the
magnetization in the film and spiral. 

To simulate the spiral rotation and obtain the preliminary estimates
of the effect, we consider the current induced spiral motion that
in turn leads to the energy flows due to the magnetic texture dynamics.
The static texture corrections due to $\eta-$ type terms will be
ignored. We consider a ferromagnetic wire with the DM interaction
in the absence of the temperature gradients. The {}``effective field''
can be found from the Free energy:\cite{Landau:1984}\begin{equation}
F=\int d^{3}\mathbf{r}\left[\dfrac{J}{2}(\boldsymbol{\nabla}\mathbf{m})^{2}+\Gamma\mathbf{m}\cdot(\boldsymbol{\nabla}\times\mathbf{m})\right]\,,\label{DM_Free_Energy}\end{equation}
where $J$ is the exchange coupling constant and $\Gamma$ is the
strength of the DM interaction. The {}``effective field'' $\mathbf{H}\equiv\partial_{\mathbf{m}}F$
can be used in the LLG Eq. (\ref{LLGD-heat1}). The ground state of
the Free energy in Eq. (\ref{DM_Free_Energy}) is a spiral state $\mathbf{m}(\mathbf{r})=\mathbf{n}_{1}\cos\mathbf{k}\cdot\mathbf{r}+\mathbf{n}_{2}\sin\mathbf{k}\cdot\mathbf{r}$
where the wavevector $\mathbf{k}=\mathbf{n}_{3}\Gamma/J$, and $\mathbf{n}_{i}$
form the right handed orthonormal vector sets. We assume that the
wavevector is along the wire, \textit{e.g.} due to anisotropies. As
can be found from the LLG Eq. (\ref{LLGD-heat}), for the case of
vanishing temperature gradients, the spiral starts to move along the
wire in the presence of currents and the solution can be described
as\cite{Goto:jul2008} \[
\mathbf{m}(\mathbf{r},t)=m_{x}\mathbf{x}+m_{\perp}\left(\mathbf{y}\cos\left[k(x-\upsilon t)\right]+\mathbf{z}\sin\left[k(x-\upsilon t)\right]\right),\]
where the $x$ axis points along the wire axis, $m_{x}=(j\hbar/e)(\beta/\alpha-1)/(2\Gamma-Jk)$,
$m_{\perp}=\sqrt{1-m_{x}^{2}}$ and $\upsilon=\wp j(\tilde{\beta}/\alpha)s\hbar/(2e)$.
The wavenumber $k=2\pi/\lambda$ and $m_{x}$ have been calculated
numerically in Ref. \onlinecite{Goto:jul2008} and in the presence
of currents $\lambda$ increases and $m_{x}$ acquires some finite
value; however, for an estimate corresponding to moderate currents
the values given by the static spiral $k_{0}=\Gamma/J$ should suffice.
The maximum current that the spiral can sustain without breaking into
the chaotic motion is $j_{max}\sim2\Gamma e/\hbar$.\cite{Goto:jul2008} 

Using Eq. (\ref{HeatCurr1}), we are now ready to find the energy
flow accompanying the current flow as the spiral moves with the speed
$\upsilon$ in the absence of the temperature gradients: $j_{q}=\Pi_{0}j-m_{\perp}^{2}k^{2}\upsilon p'\beta'/\zeta\approx0.8\Pi_{0}j$,
where only $\beta'-$ term contributes to the energy flow in Eq. (\ref{HeatCurr1}).
For our estimate, we take parameters corresponding to a MnSi:\cite{Muhlbauer:2009}
the lattice constant $a=0.5$ nm, the magnetization density $M=0.4\mu_{B}/a^{3}$,
$\lambda=20$ nm, $\alpha=0.01$, $\tilde{\beta}=\beta'=0.03$, $\sigma=5\times10^{7}\;\Omega^{-1}\mbox{m}^{-1}$,
$Ja=0.02\;\mbox{eV}$ and $\wp=\wp_{S}=0.8$. By increasing $\tilde{\beta}(\beta')$
and $\sigma$ and diminishing $\lambda$ and $\alpha$, the energy
flow can be made larger. We conclude then that the current-induced
magnetic texture dynamics can lead to additional energy flows that
in some cases can be comparable to the energy flows due to the Peltier
effect. The renormalization of the Peltier coefficient should also
apply to the Seebeck coefficient due to the Onsager reciprocity principle
dictating that $S=\Pi/T$. Assuming the absence of the temperature
gradient, from Eq. (\ref{ohmD}), we also find correction to the conductivity
caused by the EMF due to the spiral motion: $-\partial_{x}\mu=j/\sigma_{0}+m_{\perp}^{2}k^{2}\upsilon\wp\tilde{\beta}\hbar/(2e)\approx0.8j/\sigma_{0}$.

Finally, we calculate the speed of the spiral motion induced by temperature
gradients in the absence of charge currents. In full analogy to the
spiral motion induced by the charge currents and using Eq. (\ref{LLGD-heat1}),
we find the spiral speed induced by the temperature gradients:\begin{equation}
\upsilon=p'(-\kappa\partial_{x}T)\dfrac{\beta'}{\alpha}\dfrac{s\hbar}{2e}=\dfrac{\hbar}{2e}\wp_{S}S\partial_{x}T\sigma(1-\wp^{2})\dfrac{\beta'}{\alpha}\dfrac{\hbar M}{2e\gamma}\,.\label{Spiral_Speed}\end{equation}
Continuing this analogy between the energy currents and charge currents,
we can generalize the applicability of the result in Eq. (\ref{Spiral_Speed})
to transverse Neel DW\cite{Tserkovnyak:apr2008a} under the assumption
of constant temperature gradients and vanishing charge currents. Just
like the $\beta$ term is important for the current-driven DW dynamics,\cite{Tserkovnyak:apr2008a}
the new viscous $\beta'$ term is important for the thermally induced
DW motion below the Walker breakdown. 

To conclude, we phenomenologically introduced new $\mbox{\ensuremath{\beta}}-$like
viscous term into the LLG equation for the energy currents. We speculate
on a possibility of creating heat flows by microwave-induced periodic
magnetization dynamics which should result in effective cooling of
some regions, in analogy to the Peltier effect. To support it, we
considered the DM spiral texture subject to charge current and found
that the texture-dynamics induced heat flow is proportional to the
Peltier coefficient and the new viscous coupling constant $\beta'$.
Thus, the materials with large Peltier coefficient and large $\beta'$
should be suitable for the realization of the microwave cooling by
magnetization texture dynamics. The effective Peltier/Seebeck coefficient
as well as the conductivity can be modified and tuned by the texture
dynamics. These effects should be measurable in magnetic textures
characterized by the length as small as $\sim10$ nm and large viscous
damping $\beta$. Even in situations of pinned textures, the effects
of EMF induced by viscous $\beta-$like term should be seen in measurements
of the ac conductivity. We also conclude that the new viscous $\beta'$
term is important for the thermally induced DW motion. Bauer \textit{et
al.}\cite{Bauer:unpublished} worked out very similar ideas for cooling
by DW motion and thermoelectric excitation of magnetization dynamics.

We acknowledge the stimulating discussions with G. E. W. Bauer, A.
Brataas and C. H. Wong. This work was supported in part by the Alfred
P. Sloan Foundation.

\bibliographystyle{apsrev}
\bibliography{Berryheat}

\begin{thebibliography}{29}
\expandafter\ifx\csname natexlab\endcsname\relax\def\natexlab#1{#1}\fi
\expandafter\ifx\csname bibnamefont\endcsname\relax
  \def\bibnamefont#1{#1}\fi
\expandafter\ifx\csname bibfnamefont\endcsname\relax
  \def\bibfnamefont#1{#1}\fi
\expandafter\ifx\csname citenamefont\endcsname\relax
  \def\citenamefont#1{#1}\fi
\expandafter\ifx\csname url\endcsname\relax
  \def\url#1{\texttt{#1}}\fi
\expandafter\ifx\csname urlprefix\endcsname\relax\def\urlprefix{URL }\fi
\providecommand{\bibinfo}[2]{#2}
\providecommand{\eprint}[2][]{\url{#2}}

\bibitem[{\citenamefont{Berry}(1984)}]{Berry:mar1984}
\bibinfo{author}{\bibfnamefont{M.~V.} \bibnamefont{Berry}},
  \bibinfo{journal}{Proc. R. Soc. London, Ser. A}
  \textbf{\bibinfo{volume}{392}}, \bibinfo{pages}{45} (\bibinfo{year}{1984}).

\bibitem[{\citenamefont{Barnes and Maekawa}(2007)}]{Barnes:jun2007}
\bibinfo{author}{\bibfnamefont{S.~E.} \bibnamefont{Barnes}} \bibnamefont{and}
  \bibinfo{author}{\bibfnamefont{S.}~\bibnamefont{Maekawa}},
  \bibinfo{journal}{Phys. Rev. Lett.} \textbf{\bibinfo{volume}{98}},
  \bibinfo{eid}{246601} (\bibinfo{year}{2007}).

\bibitem[{\citenamefont{{Tserkovnyak} and
  {Mecklenburg}}(2008)}]{Tserkovnyak:apr2008}
\bibinfo{author}{\bibfnamefont{Y.}~\bibnamefont{{Tserkovnyak}}}
  \bibnamefont{and}
  \bibinfo{author}{\bibfnamefont{M.}~\bibnamefont{{Mecklenburg}}},
  \bibinfo{journal}{Phys. Rev. B} \textbf{\bibinfo{volume}{77}},
  \bibinfo{pages}{134407} (\bibinfo{year}{2008}).

\bibitem[{\citenamefont{{Volovik}}(1987)}]{Volovik:mar1987}
\bibinfo{author}{\bibfnamefont{G.~E.} \bibnamefont{{Volovik}}},
  \bibinfo{journal}{J. Phys. C} \textbf{\bibinfo{volume}{20}},
  \bibinfo{pages}{L83} (\bibinfo{year}{1987}).

\bibitem[{\citenamefont{{Tserkovnyak} and {Wong}}(2009)}]{Tserkovnyak:jan2009}
\bibinfo{author}{\bibfnamefont{Y.}~\bibnamefont{{Tserkovnyak}}}
  \bibnamefont{and} \bibinfo{author}{\bibfnamefont{C.~H.}
  \bibnamefont{{Wong}}}, \bibinfo{journal}{Phys. Rev. B}
  \textbf{\bibinfo{volume}{79}}, \bibinfo{pages}{014402}
  (\bibinfo{year}{2009}).

\bibitem[{\citenamefont{{Zhang} and {Li}}(2004)}]{Zhang:sep2004}
\bibinfo{author}{\bibfnamefont{S.}~\bibnamefont{{Zhang}}} \bibnamefont{and}
  \bibinfo{author}{\bibfnamefont{Z.}~\bibnamefont{{Li}}},
  \bibinfo{journal}{Phys. Rev. Lett.} \textbf{\bibinfo{volume}{93}},
  \bibinfo{pages}{127204} (\bibinfo{year}{2004});
\bibinfo{author}{\bibfnamefont{A.}~\bibnamefont{Thiaville}},
  \bibinfo{author}{\bibfnamefont{Y.}~\bibnamefont{Nakatani}},
  \bibinfo{author}{\bibfnamefont{J.}~\bibnamefont{Miltat}}, \bibnamefont{and}
  \bibinfo{author}{\bibfnamefont{Y.}~\bibnamefont{Suzuki}},
  \bibinfo{journal}{Europhys. Lett.} \textbf{\bibinfo{volume}{69}},
  \bibinfo{pages}{990} (\bibinfo{year}{2005});
\bibinfo{author}{\bibfnamefont{H.}~\bibnamefont{Kohno}},
  \bibinfo{author}{\bibfnamefont{G.}~\bibnamefont{Tatara}}, \bibnamefont{and}
  \bibinfo{author}{\bibfnamefont{J.}~\bibnamefont{Shibata}},
  \bibinfo{journal}{J. Phys. Soc. Jpn.} \textbf{\bibinfo{volume}{75}},
  \bibinfo{pages}{113706} (\bibinfo{year}{2006});
\bibinfo{author}{\bibfnamefont{R.~A.} \bibnamefont{Duine}},
  \bibinfo{author}{\bibfnamefont{A.~S.} \bibnamefont{Nunez}},
  \bibinfo{author}{\bibfnamefont{J.}~\bibnamefont{Sinova}}, \bibnamefont{and}
  \bibinfo{author}{\bibfnamefont{A.~H.} \bibnamefont{MacDonald}},
  \bibinfo{journal}{Phys. Rev. B} \textbf{\bibinfo{volume}{75}},
  \bibinfo{eid}{214420} (\bibinfo{year}{2007}).

\bibitem[{\citenamefont{Tserkovnyak et~al.}(2008)\citenamefont{Tserkovnyak,
  Brataas, and Bauer}}]{Tserkovnyak:apr2008a}
\bibinfo{author}{\bibfnamefont{Y.}~\bibnamefont{Tserkovnyak}},
  \bibinfo{author}{\bibfnamefont{A.}~\bibnamefont{Brataas}}, \bibnamefont{and}
  \bibinfo{author}{\bibfnamefont{G.~E.} \bibnamefont{Bauer}},
  \bibinfo{journal}{J. Magn. Magn. Mater.} \textbf{\bibinfo{volume}{320}},
  \bibinfo{pages}{1282 } (\bibinfo{year}{2008}).

\bibitem[{\citenamefont{Sales}(2002)}]{Sales:feb2002}
\bibinfo{author}{\bibfnamefont{B.~C.} \bibnamefont{Sales}},
  \bibinfo{journal}{Science} \textbf{\bibinfo{volume}{295}},
  \bibinfo{pages}{1248} (\bibinfo{year}{2002});
\bibinfo{author}{\bibfnamefont{A.~I.} \bibnamefont{Hochbaum}},
  \bibinfo{author}{\bibfnamefont{R.}~\bibnamefont{Chen}},
  \bibinfo{author}{\bibfnamefont{R.~D.} \bibnamefont{Delgado}},
  \bibinfo{author}{\bibfnamefont{W.}~\bibnamefont{Liang}},
  \bibinfo{author}{\bibfnamefont{E.~C.} \bibnamefont{Garnett}},
  \bibinfo{author}{\bibfnamefont{M.}~\bibnamefont{Najarian}},
  \bibinfo{author}{\bibfnamefont{A.}~\bibnamefont{Majumdar}}, \bibnamefont{and}
  \bibinfo{author}{\bibfnamefont{P.}~\bibnamefont{Yang}},
  \bibinfo{journal}{Nature} \textbf{\bibinfo{volume}{451}},
  \bibinfo{pages}{163} (\bibinfo{year}{2008}),
\bibinfo{author}{\bibfnamefont{A.~I.} \bibnamefont{Boukai}},
  \bibinfo{author}{\bibfnamefont{Y.}~\bibnamefont{Bunimovich}},
  \bibinfo{author}{\bibfnamefont{J.}~\bibnamefont{Tahir-Kheli}},
  \bibinfo{author}{\bibfnamefont{J.-K.} \bibnamefont{Yu}},
  \bibinfo{author}{\bibfnamefont{W.~A.} \bibnamefont{Goddard}},
  \bibnamefont{and} \bibinfo{author}{\bibfnamefont{J.~R.} \bibnamefont{Heath}},
  \textit{ibid.} \textbf{\bibinfo{volume}{451}},
  \bibinfo{pages}{168} (\bibinfo{year}{2008}).

\bibitem[{\citenamefont{Hatami et~al.}(2009)\citenamefont{Hatami, Bauer, Zhang,
  and Kelly}}]{Hatami:may2009}
\bibinfo{author}{\bibfnamefont{M.}~\bibnamefont{Hatami}},
  \bibinfo{author}{\bibfnamefont{G.~E.~W.} \bibnamefont{Bauer}},
  \bibinfo{author}{\bibfnamefont{Q.}~\bibnamefont{Zhang}}, \bibnamefont{and}
  \bibinfo{author}{\bibfnamefont{P.~J.} \bibnamefont{Kelly}},
  \bibinfo{journal}{Phys. Rev. B} \textbf{\bibinfo{volume}{79}},
  \bibinfo{eid}{174426} (\bibinfo{year}{2009});
\bibinfo{author}{\bibfnamefont{M.}~\bibnamefont{{Hatami}}},
  \bibinfo{author}{\bibfnamefont{G.~E.~W.} \bibnamefont{{Bauer}}},
  \bibinfo{author}{\bibfnamefont{Q.}~\bibnamefont{{Zhang}}}, \bibnamefont{and}
  \bibinfo{author}{\bibfnamefont{P.~J.} \bibnamefont{{Kelly}}},
  \bibinfo{journal}{Phys. Rev. Lett.} \textbf{\bibinfo{volume}{99}},
  \bibinfo{pages}{066603} (\bibinfo{year}{2007}).

\bibitem[{\citenamefont{{Berger}}(1985)}]{Berger:jul1985}
\bibinfo{author}{\bibfnamefont{L.}~\bibnamefont{{Berger}}},
  \bibinfo{journal}{J. Appl. Phys.} \textbf{\bibinfo{volume}{58}},
  \bibinfo{pages}{450} (\bibinfo{year}{1985}),
\bibinfo{author}{\bibfnamefont{S.~U.} \bibnamefont{{Jen}}} \bibnamefont{and}
  \bibinfo{author}{\bibfnamefont{L.}~\bibnamefont{{Berger}}},
  \textit{ibid.} \textbf{\bibinfo{volume}{59}},
  \bibinfo{pages}{1278} (\bibinfo{year}{1986}{\natexlab{a}}),
\bibinfo{author}{\bibfnamefont{S.~U.} \bibnamefont{{Jen}}} \bibnamefont{and}
  \bibinfo{author}{\bibfnamefont{L.}~\bibnamefont{{Berger}}},
  \textit{ibid.} \textbf{\bibinfo{volume}{59}},
  \bibinfo{pages}{1285} (\bibinfo{year}{1986}{\natexlab{b}}).

\bibitem[{\citenamefont{Ohta et~al.}(2007)\citenamefont{Ohta, Kim, Mune,
  Mizoguchi, Nomura, Ohta, Nomura, Nakanishi, Ikuhara, Hirano
  et~al.}}]{Ohta:feb2007}
\bibinfo{author}{\bibfnamefont{H.}~\bibnamefont{Ohta}},
  \bibinfo{author}{\bibfnamefont{S.}~\bibnamefont{Kim}},
  \bibinfo{author}{\bibfnamefont{Y.}~\bibnamefont{Mune}},
  \bibinfo{author}{\bibfnamefont{T.}~\bibnamefont{Mizoguchi}},
  \bibinfo{author}{\bibfnamefont{K.}~\bibnamefont{Nomura}},
  \bibinfo{author}{\bibfnamefont{S.}~\bibnamefont{Ohta}},
  \bibinfo{author}{\bibfnamefont{T.}~\bibnamefont{Nomura}},
  \bibinfo{author}{\bibfnamefont{Y.}~\bibnamefont{Nakanishi}},
  \bibinfo{author}{\bibfnamefont{Y.}~\bibnamefont{Ikuhara}},
  \bibinfo{author}{\bibfnamefont{M.}~\bibnamefont{Hirano}},
  \bibnamefont{et~al.}, \bibinfo{journal}{Nature Mater.}
  \textbf{\bibinfo{volume}{6}}, \bibinfo{pages}{129} (\bibinfo{year}{2007}).

\bibitem[{\citenamefont{Yamaguchi et~al.}(2004)\citenamefont{Yamaguchi, Ono,
  Nasu, Miyake, Mibu, and Shinjo}}]{Yamaguchi:Feb2004}
\bibinfo{author}{\bibfnamefont{A.}~\bibnamefont{Yamaguchi}},
  \bibinfo{author}{\bibfnamefont{T.}~\bibnamefont{Ono}},
  \bibinfo{author}{\bibfnamefont{S.}~\bibnamefont{Nasu}},
  \bibinfo{author}{\bibfnamefont{K.}~\bibnamefont{Miyake}},
  \bibinfo{author}{\bibfnamefont{K.}~\bibnamefont{Mibu}}, \bibnamefont{and}
  \bibinfo{author}{\bibfnamefont{T.}~\bibnamefont{Shinjo}},
  \bibinfo{journal}{Phys. Rev. Lett.} \textbf{\bibinfo{volume}{92}},
  \bibinfo{pages}{077205} (\bibinfo{year}{2004}),
\bibinfo{author}{\bibfnamefont{M.}~\bibnamefont{Hayashi}},
  \bibinfo{author}{\bibfnamefont{L.}~\bibnamefont{Thomas}},
  \bibinfo{author}{\bibfnamefont{Y.~B.} \bibnamefont{Bazaliy}},
  \bibinfo{author}{\bibfnamefont{C.}~\bibnamefont{Rettner}},
  \bibinfo{author}{\bibfnamefont{R.}~\bibnamefont{Moriya}},
  \bibinfo{author}{\bibfnamefont{X.}~\bibnamefont{Jiang}}, \bibnamefont{and}
  \bibinfo{author}{\bibfnamefont{S.~S.~P.} \bibnamefont{Parkin}},
  \textit{ibid.} \textbf{\bibinfo{volume}{96}},
  \bibinfo{eid}{197207} (\bibinfo{year}{2006});
\bibinfo{author}{\bibfnamefont{M.}~\bibnamefont{Hayashi}},
  \bibinfo{author}{\bibfnamefont{L.}~\bibnamefont{Thomas}},
  \bibinfo{author}{\bibfnamefont{C.}~\bibnamefont{Rettner}},
  \bibinfo{author}{\bibfnamefont{R.}~\bibnamefont{Moriya}}, \bibnamefont{and}
  \bibinfo{author}{\bibfnamefont{S.~S.~P.} \bibnamefont{Parkin}},
  \bibinfo{journal}{Nat. Phys.} \textbf{\bibinfo{volume}{3}},
  \bibinfo{pages}{21} (\bibinfo{year}{2007}).

\bibitem[{\citenamefont{Muhlbauer et~al.}(2009)\citenamefont{Muhlbauer, Binz,
  Jonietz, Pfleiderer, Rosch, Neubauer, Georgii, and Boni}}]{Muhlbauer:2009}
\bibinfo{author}{\bibfnamefont{S.}~\bibnamefont{Muhlbauer}},
  \bibinfo{author}{\bibfnamefont{B.}~\bibnamefont{Binz}},
  \bibinfo{author}{\bibfnamefont{F.}~\bibnamefont{Jonietz}},
  \bibinfo{author}{\bibfnamefont{C.}~\bibnamefont{Pfleiderer}},
  \bibinfo{author}{\bibfnamefont{A.}~\bibnamefont{Rosch}},
  \bibinfo{author}{\bibfnamefont{A.}~\bibnamefont{Neubauer}},
  \bibinfo{author}{\bibfnamefont{R.}~\bibnamefont{Georgii}}, \bibnamefont{and}
  \bibinfo{author}{\bibfnamefont{P.}~\bibnamefont{Boni}},
  \bibinfo{journal}{Science} \textbf{\bibinfo{volume}{323}},
  \bibinfo{pages}{915} (\bibinfo{year}{2009}).

\bibitem[{\citenamefont{Landau and Lifshitz}(1984)}]{Landau:1984}
\bibinfo{author}{\bibfnamefont{L.}~\bibnamefont{Landau}} \bibnamefont{and}
  \bibinfo{author}{\bibfnamefont{E.}~\bibnamefont{Lifshitz}},
  \emph{\bibinfo{title}{Electrodynamics of Continuous Media}},
  vol.~\bibinfo{volume}{8} (\bibinfo{publisher}{Pergamon, Oxford},
  \bibinfo{year}{1984}), \bibinfo{edition}{2nd} ed.

\bibitem[{\citenamefont{Gilbert}(2004)}]{Gilbert:nov2004}
\bibinfo{author}{\bibfnamefont{T.}~\bibnamefont{Gilbert}},
  \bibinfo{journal}{IEEE Trans. Magn.} \textbf{\bibinfo{volume}{40}},
  \bibinfo{pages}{3443} (\bibinfo{year}{2004}).

\bibitem[{\citenamefont{{D'amico} and {Vignale}}(2000)}]{D'amico:aug2000}
\bibinfo{author}{\bibfnamefont{I.}~\bibnamefont{{D'amico}}} \bibnamefont{and}
  \bibinfo{author}{\bibfnamefont{G.}~\bibnamefont{{Vignale}}},
  \bibinfo{journal}{Phys. Rev. B} \textbf{\bibinfo{volume}{62}},
  \bibinfo{pages}{4853} (\bibinfo{year}{2000}).

\bibitem[{\citenamefont{Uzdin and Vega}(2008)}]{Uzdin:jun2008}
\bibinfo{author}{\bibfnamefont{V.~M.} \bibnamefont{Uzdin}} \bibnamefont{and}
  \bibinfo{author}{\bibfnamefont{A.}~\bibnamefont{Vega}},
  \bibinfo{journal}{Nanotechnology} \textbf{\bibinfo{volume}{19}},
  \bibinfo{pages}{315401} (\bibinfo{year}{2008}).

\bibitem[{\citenamefont{{Goto} et~al.}(unpublished)\citenamefont{{Goto},
  {Katsura}, and {Nagaosa}}}]{Goto:jul2008}
\bibinfo{author}{\bibfnamefont{K.}~\bibnamefont{{Goto}}},
  \bibinfo{author}{\bibfnamefont{H.}~\bibnamefont{{Katsura}}},
  \bibnamefont{and}
  \bibinfo{author}{\bibfnamefont{N.}~\bibnamefont{{Nagaosa}}},
  \bibinfo{journal}{arXiv:0807.2901}  (\bibinfo{year}{unpublished}).

\bibitem[{\citenamefont{{Bauer} et~al.}(unpublished)\citenamefont{{Bauer},
  {Bretzel}, {Hatami}, and {Brataas}}}]{Bauer:unpublished}
\bibinfo{author}{\bibfnamefont{G.~E.~W.} \bibnamefont{{Bauer}}}
  \textit{et al.}, 
  \bibinfo{journal}{preprint}  (\bibinfo{year}{unpublished}).

\end{thebibliography}

\end{document}